\documentclass[12pt]{article}
\usepackage{amsmath}
\usepackage{graphicx}
\usepackage{enumerate}
\usepackage{natbib}
\usepackage{amssymb}

\usepackage{url} 
\newcommand{\blind}{1}

\usepackage{graphicx}

\usepackage{multirow}
\usepackage{comment}
\usepackage{amsmath, amssymb, amsfonts, bbm}
\usepackage{mathrsfs}

\DeclareMathOperator*{\argmin}{arg\,min}

\newcommand{\pr}{P} 
\newcommand{\tr}{{\rm{tr}}}
\newcommand{\logit}{\mbox{logit}} 
\newcommand{\ba}{\boldsymbol{a}}
\newcommand{\bA}{\boldsymbol{A}}

\newcommand{\bd}{\boldsymbol{d}}

\newcommand{\bI}{\boldsymbol{I}}

\newcommand{\bx}{\boldsymbol{x}}
\newcommand{\bX}{\boldsymbol{X}}

\newcommand{\by}{\boldsymbol{y}}

\newcommand{\bz}{\boldsymbol{z}}

\newcommand{\balpha}{\boldsymbol{\alpha}}

\newcommand{\var}{\mathrm{Var}}

\def\trans{^{\rm T}}

\newcommand{\bS}{\boldsymbol{S}}

\newcommand{\bK}{\mathbf{K}}
\newcommand{\bphi}{\boldsymbol{\phi}}

\newcommand{\bone}{\boldsymbol{1}}
\newcommand{\bbeta}{\boldsymbol{\beta}}

\newcommand{\bvar}{\boldsymbol{\varepsilon}}

\newcommand{\bDelta}{\boldsymbol{\Delta}}

\newcommand{\bPi}{\boldsymbol{\Pi}}

\newcommand{\Var}{\mbox{Var}}

\newtheorem{theorem}{Theorem}

\newtheorem{remark}{Remark}

\newtheorem{lemma}{Lemma}

\def\wt{\widetilde}
\newcommand{\Norm}[1]{\left\Vert#1\right\Vert}
\newcommand{\norm}[1]{\left\Vert#1\right\Vert}

\usepackage{xcolor}

\usepackage{multirow,xcolor}

\usepackage{xr}
\makeatletter
\newcommand*{\addFileDependency}[1]{
  \typeout{(#1)}
  \@addtofilelist{#1}
  \IfFileExists{#1}{}{\typeout{No file #1.}}
}
\makeatother

\newcommand*{\myexternaldocument}[1]{%
    \externaldocument{#1}%
    \addFileDependency{#1.tex}%
    \addFileDependency{#1.aux}%
}
\myexternaldocument{supplement}

\def\spacingset#1{\renewcommand{\baselinestretch}%
{#1}\small\normalsize} \spacingset{1}

\begin{document} 

\if1\blind
{
  \title{\bf 
  Statistical inference using  Regularized M-estimation in the reproducing kernel Hilbert space for handling  missing data 
  }
  \author{Hangfang Wang 
    \and 
    Jae Kwang Kim \thanks{Department of Statistics, Iowa State University, Ames, IA 50011, U.S.A.}}
  \maketitle
} \fi

\if0\blind
{
  \bigskip
  \bigskip
  \bigskip
  \begin{center}
    {\LARGE\bf 
    Statistical inference using  Regularized M-estimation in the reproducing kernel Hilbert space for handling  missing data
    }
\end{center}
  \medskip
} \fi

\bigskip
\begin{abstract}
Imputation and propensity score weighting are two popular techniques for handling missing data. We address these problems using the regularized M-estimation  techniques in the reproducing kernel Hilbert space. Specifically, we first use the kernel ridge  regression to develop imputation for handling item nonresponse. While this nonparametric approach is potentially promising for imputation, its statistical properties are not investigated in the literature. Under some conditions on the order of the tuning parameter, we first establish the  root-$n$ consistency of the kernel ridge regression imputation estimator and show that it  achieves the lower bound of the semiparametric asymptotic variance.  A nonparametric  propensity score estimator using the reproducing kernel Hilbert space is also developed by a novel application of the maximum entropy method for the density ratio function estimation. We show that the resulting propensity score estimator is asymptotically equivalent to the kernel ridge regression imputation estimator. 
Results from a limited simulation study are also presented to confirm our theory. The proposed method is applied to analyze the air pollution data measured in Beijing,  China.

\end{abstract}

\noindent%
\vfill

\newpage
\spacingset{1.5} 

\maketitle

\section{Introduction}

Missing data is a universal problem in statistics.  Ignoring the cases with missing values  can lead to  misleading results \citep{kim2013statistical, little2019statistical}. 
Two popular approaches for handling missing data are imputation and propensity score weighting. Both approaches are based on some assumptions about the data structure and the response mechanism. To avoid potential biases due to model misspecification, instead of using strong parametric model assumptions, nonparametric approaches are preferred as they do not depend  on explicit model assumptions. 

In principle, any prediction techniques can be used to impute for   missing values using the responding units as a training sample. However, statistical inference with  imputed estimator is not straightforward. Treating imputed data as if observed and applying the standard estimation procedure may result in misleading inference, leading to  underestimation of the variance  of  imputed point estimators.  
How to incorporate the uncertainty of the estimated parameters in the final inference is challenging especially for nonparametric imputation because the model parameter is implicitly defined. 

For  nonparametric imputation, \citet{cheng1994nonparametric} used  the kernel-based nonparametric regression for imputation and established the root-$n$ consistency of the imputed estimator. \cite{chen01} considered nearest neighbor imputation and discuss its variance estimation. 
\citet{wang09} employed the kernel smoothing approach to do empirical likelihood inference with missing values. \cite{yang2020b} considered predictive mean matching imputation and established its asymptotic properties. 
\cite{kim2014multiple}  proposed Bayesian multiple imputation using the Dirichlet process mixture. 
\cite{sang2020} proposed semiparametric fractional imputation using Gaussian mixtures.

For  nonparametric propensity score estimation, 
\cite{hainmueller2012} proposed so-called the entropy balancing method to find the propensity score weights using the Kullback-Leibler information criterion with finite-dimensional basis function. \cite{chen2013} established the root-${n}$ consistency of the kernel-based nonparametric propensity score estimator. 
\cite{chan2016} generalized the entropy balancing method of \cite{hainmueller2012} further to develop a general calibration weighting method that satisfies the covariance balancing property with increasing dimensions of the control variables. They further showed the global efficiency of the proposed calibration weighting estimator.  \cite{zhao2019} generalized the idea further and developed a unified approach of covariate balancing propensity score  method using tailored loss functions. \cite{tan2020regularized} developed regularized calibrated estimation of propensity scores with high dimensional covariates.  While nonparametric kernel regression can be used to construct nonparametric propensity score estimation, as in \cite{chen2013}, it is not clear how to generalize it to a wider function space to obtain nonparametric propensity score estimation. 


In this paper, we consider regularized M-estimation  as a tool for nonparametric function estimation for imputation and  propensity score estimation. Kernel ridge regression \citep{friedman2001elements,shawe2004kernel} is an example of the regularized M-estimation for a modern regression technique. {By using} a regularized M-estimator {in reproducing} kernel Hilbert space (RKHS), kernel ridge regression can estimate the regression mean function with {complex reproducing kernel Hilbert space} while a regularized term makes the original infinite dimensional estimation problem viable \citep{wahba1990spline}. Due to its flexibility in the choice of kernel functions, 
kernel ridge regression is very popular in machine learning. 
\citet{geer2000empirical,mendelson2002geometric,zhang2005learning,koltchinskii2006local,steinwart2009optimal} studied the error bounds for the estimates of kernel ridge regression method.

While  the kernel ridge regression is a promising tool for handling missing data, its statistical inference   is not   investigated in the literature. We aim to fill in this important research gap in the missing data literature by establishing the statistical properties of the KRR imputation estimator. 
  Specifically, we obtain  root-$n$ consistency of the KRR  imputation estimator under some popular functional Hilbert spaces. 
Because the KRR  is a general tool for nonparametric regression with flexible assumptions, the proposed imputation method can be used widely to handle missing data without employing parameteric model assumptions. Variance estimation after the kernel ridge regression imputation is a challenging but important problem.  To the best of our knowledge, this is the first paper which considers kernel ridge regression technique for imputation and discusses its variance estimation rigorously.

The regularized M-estimation technique in RKHS is also used to obtain nonparametric propensity score weights for handling missing data. To do this, we use a novel application of density ratio function estimation in the same reproducing kernel Hilbert space. Maximum entropy method of \citet{nguyen2010} for density ratio estimation is adopted to get the nonparametric  propensity score estimators. We further show the asymptotic equivalence of the resulting propensity score estimator with the kernel ridge regression-based imputation estimator. These theoretical findings can be used to make valid statistical inferences with the propensity score estimator.

The paper is organized as follows. In Section 2, the basic setup and the KRR  method is introduced. In Section 3,    the root-$n$ consistency of the KRR imputation estimator  is established. In Section 4, we introduce a novel  nonparametric propensity score  estimator using the regularized M-estimation technique in the RKHS.   Results from a limited simulation study are presented in Section 5. An illustration of the proposed method to a real data example is presented in Section 6. Some concluding remarks are made in Section 7.

\section{Basic setup}

Consider the problem of estimating $\theta=E(Y)$ from an independent and identically distributed  sample  $\{(\bx_i, y_i), i=1, \cdots, n\}$ of random {vector} $(\bX,Y)$. Instead of always {observing  $y_i$}, suppose that we observe $y_i$ only if $\delta_i=1$, where $\delta_i$ is the response indicator function of unit $i$ taking values on $\{0,1\}$.  The auxiliary variable $\bx_i$ are always observed. 
We assume that the response mechanism is missing at random  (MAR) in the sense of   \cite{rubin1976}. 

Under MAR, we can develop a nonparametric estimator $\hat{m} (\bx)$ of $m(\bx)=E( Y \mid \bx)$ and construct the following imputation estimator: 
\begin{equation} 
\hat{\theta}_I = \frac{1}{n} \sum_{i=1}^n \left\{ \delta_i y_i + (1-\delta_i )\hat{m} (\bx_i) \right\}. 
\label{1} 
\end{equation} 
If $\hat{m} (\bx)$ is constructed by the kernel-based nonparametric regression method, we can express 
\begin{equation} 
 \hat{m} (\bx) = \frac{ \sum_{i=1}^n \delta_i K_h( \bx_i, \bx ) y_i}{ \sum_{i=1}^n \delta_i K_h( \bx_i, \bx )} 
\label{2}
\end{equation} 
where $K_h (\cdot)$ is the kernel function with bandwidth $h$. Under some suitable choice of the bandwidth $h$, \citet{cheng1994nonparametric} first established the root-$n$ consistency of the imputation estimator (\ref{1}) with nonparametric  function in (\ref{2}).  However, the kernel-based regression imputation in (\ref{2}) is applicable only when the dimension of $\bx$ is small.

%

In this paper, we extend the work of \citet{cheng1994nonparametric} by considering a more general type of the nonparametric imputation, called kernel ridge regression  imputation. The kernel ridge regression (KRR) can be understood using the reproducing kernel Hilbert space theory \citep{aronszajn1950theory} and can be described as 
\begin{equation} 
\hat{m} = \argmin_{m\in \mathcal{H}} \left[  \sum_{i=1}^{n} \delta_{i}\left\{ y_{i} - m(\bx_{i}) \right\}^{2} + \lambda \Norm{m}_{\mathcal{H}}^{2}   \right],
\label{3} 
\end{equation} 
where $ \norm{m}_{\mathcal{H}}^{2}  $ is the norm of $m$ in the reproducing kernel Hilbert space $\mathcal{H}$ and $\lambda(>0)$ is a tuning parameter for regularization. Here, the inner product $\langle \cdot, \cdot \rangle_{\mathcal{H}}$ is induced by {such} a kernel function, i.e., 
\begin{align}
\langle f, K(\cdot, \bx) \rangle_{\mathcal{H}} = f(\bx), \notag
\end{align}
for any  $\bx \in \mathcal{X},  f\in \mathcal{H}$,
namely, the reproducing property of $\mathcal{H}$. Naturally, this  reproducing property implies the $\mathcal{H}$ norm of $f$: $\norm{f}_{\mathcal{H}} = \langle f, f  \rangle_{\mathcal{H}}^{1/2}$.
\cite{scholkopf2002} provides a comprehensive overview of the machine learning techniques using the reproducing kernel functions. 

One 
canonical example of such a functional Hilbert space is the Sobolev space. Specifically, assuming that  the domain of such functional space is $[0,1]$,
 the Sobolev space of order $\ell$ can be denoted as 
\begin{eqnarray}
  \mathcal{W}_{2}^{\ell} &=& \left\{ f:[0,1] \rightarrow \mathbb{R}  \mid 
   f, f^{(1)}, \dots, f^{(\ell-1)} \subset \mathbb{C}[0,1], \quad f^{(\ell)} \in L^{2}[0,1]    \right\}, \notag 
\end{eqnarray}
where $\mathbb{C}[0,1]$ denotes the absolutely continuous function on $[0,1]$.
One possible norm for this space can be
\begin{eqnarray}
  \norm{f}_{\mathcal{W}_{2}^{\ell}}^{2} = \sum_{q = 0}^{\ell-1}\left\{   \int_{0}^{1}f^{(q)}(t)dt      \right\}^{2} +
   \int_{0}^{1}\left\{f^{(\ell)}(t) \right\}^{2}dt . \notag 
\end{eqnarray}
In this section, we employ the Sobolev space of second order as the approximation function space. 
For Sobolev space of order $\ell$, we have the kernel function
\begin{align}
K(x,y) = \sum_{q = 0}^{\ell-1}k_{q}(x)k_{q}(y) + k_{\ell}(x)k_{\ell}(y)
 + (-1)^{\ell}   k_{2\ell}(|x-y|),\notag 
\end{align}
where $k_{q}(x) = (q!)^{-1}B_{q}(x)$ and $B_{q}(\cdot)$ is the Bernoulli polynomial of order $q$. Smoothing spline method is a special case of the kernel ridge regression method.

By the representer theorem for reproducing kernel Hilbert space \citep{wahba1990spline}, the estimate  in (\ref{3})  lies in the linear span of $\{K(\cdot, \bx_{i}), i = 1,\ldots, n\}$.
Specifically, we have 
\begin{align}\label{KRR}
\hat{m}(\cdot) = \sum_{i=1}^{n}\hat{\alpha}_{i,\lambda}K(\cdot, \bx_{i}),
\end{align}
where 
\begin{align}
\hat{\balpha}_{\lambda} = \left(\bDelta_{n}\bK +  \lambda\bI_{n}\right)^{-1}\bDelta_n \by,\notag
\end{align}
$\bDelta_{n} = \mbox{diag}(\delta_{1}, \ldots, \delta_{n})$, $\bK = (K(\bx_{i}, \bx_{j}))_{ij} $,  $\by = (y_{1},\ldots, y_{n})\trans$ and $\bI_{n}$ is the $n\times n$ identity matrix.

The tuning parameter $\lambda$ is selected via generalized cross-validation in kernel ridge regeression, where the criterion for $\lambda$ is
\begin{align}\label{GCV}
 \mbox{GCV}(\lambda) = \frac{n^{-1}   \Norm{ \left\{\bDelta_{n} - \bA(\lambda)\right\}\by }_{2}^{2}   }{n^{-1}  \tr(\bDelta_{n} - \bA(\lambda) )  },
\end{align}
and $\bA(\lambda) = \bDelta_{n}\bK  ( \bDelta_{n} \bK +  \lambda \bI_{n}  )^{-1} \bDelta_{n}  $. The value of $\lambda$ minimizing the criterion \eqref{GCV} is used for the selected tuning parameter.

Using the kernel ridge regression (KRR) imputation in (\ref{3}), we can obtain the imputed estimator in (\ref{1}). Because $\hat{m}(\bx)$ in (\ref{KRR}) is a nonparametric regression estimator of $m(\bx)=E ( Y \mid \bx)$, we can expect that this imputation estimator in (\ref{1}) is consistent for $\theta=E(Y)$ under missing at random, as long as $\hat{m}(\bx)$ is a consistent estimator of $m(\bx)$. Surprisingly, it turns out  that the consistency of $\hat{\theta}_I$ to $\theta$ is of order $O_p(n^{-1/2})$, while the point-wise convergence rate for $\hat{m}( \bx)$ to $m(\bx)$ is slower. This is consistent with the theory of \cite{cheng1994nonparametric} for kernel-based nonparametric regression imputation. 

We aim to establish two goals: (i) find the sufficient conditions for the  root-$n$ consistency of the KRR imputation estimator and give a formal proof; (ii) find a linearization variance formula for the KRR imputation estimator.  The first part is formally presented in Theorem \ref{main theorem} in Section 3. For the second part, we employ the density ratio estimation method of \cite{nguyen2010}  to get a consistent estimator of  $\omega (\bx) = \{\pi (\bx)\}^{-1}$ in the linearized version of $\hat{\theta}_I$. Estimation of $\omega(\bx)$ will be presented in Section 4.

\section{Main Theory}

Before we develop our main theory, we first introduce Mercer's theorem.
\begin{lemma}[Mercer's theorem]\label{Mercer}
Given a continuous, symmetric, positive definite kernel function $K: \mathcal{X} \times \mathcal{X} \mapsto \mathbb{R}$. For $\bx, \bz \in \mathcal{X}$, under some regularity conditions, Mercer's theorem characterizes $K$ by the following expansion
\begin{align}
 K(\bx, \bz) = \sum_{j=1}^{\infty}\lambda_{j}\psi_{j}(\bx) \psi_{j}(\bz),\notag
\end{align}
where $\lambda_{1} \geq \lambda_{2} \geq \ldots \geq 0$ are a non-negative sequence of eigenvalues, $\{\psi_{j} \}_{j=1}^{\infty}$ is an orthonormal basis for $L^{2}(\mathbb{P})$ and $\mathbb{P}$ is the given distribution of $\bX$ on $\mathcal{X}$.
\end{lemma}

Furthermore, we make the following  assumptions.
\begin{description}
\item{[A1]} 
 \label{A1}
  For some $k \geq 2$, there is a constant $\rho < \infty$ such that $E[ \psi_{j}(X)^{2k} ] \leq \rho^{2k}$ for all 
  $j \in \mathbb{N}$, where $\{\psi_{j}\}_{j=1}^{\infty}$ are orthonormal basis by expansion from Mercer's theorem.
\item{[A2]} 
\label{A2}
  The function $m \in \mathcal{H}$, and for $\bx \in \mathcal{X}$, we have $E[\left\{ Y -  m(\bx)\right\}^{2} \mid \bx ] \leq \sigma^{2}$, {for some  $\sigma^{2} < \infty$.}
\item{[A3]} 
\label{A3} The response mechanism is missing at random. Furthermore, the propensity score $\pi(\bx)=\pr( \delta=1 \mid \bx)$ is uniformly bounded away from zero. In particular, there exists a positive constant $c > 0$ such that 
      $\pi(\bx_{i}) \geq c$, for $i = 1, \ldots, n$.
\end{description}

The first assumption is a technical assumption which controls the tail behavior of $\{\psi_{j}\}_{j=1}^{\infty}$. Assumption \ref{A2} indicates that the noises have bounded variance. Assumption \ref{A1} and Assumption \ref{A2} together aim to control the error bound of the kernel ridge regression estimate $\hat{m}$. {Furthermore}, Assumption \ref{A3} means that the support for the respondents should be the same as  the original sample support. Assumption \ref{A3} is a standard assumption for missing data analysis.

We further introduce the following lemma.  Let $\bS_{\lambda} =  (  \bI_{n} + \lambda \bK^{-1}   )^{-1}$ be the linear smoother for the KRR method. That is, $\hat{m}= \bS_{\lambda} \by$ be the vector of regression predictor of $\by$ using the kernel ridge regression method. We now present the following lemma without proof which is modified from Lemma 7 in \citet{zhang2013divide}.

\begin{lemma}\label{order}
Under [A1]-[A2], for a random vector $\bz = E(\bz) + \sigma \bvar$, we have
\begin{align}
 \bS_{\lambda}\bz = E(\bz \mid \bx) + \ba_n,\notag
\end{align}
where $\ba_n = (a_{1}, \ldots, a_{n})\trans$ and
\begin{align}
a_{i} = \mathcal{O}_{p} \left( \lambda^{1/2}   +  \{\gamma(\lambda)\}^{1/2}n^{-1/2}  \right),
\label{a1}
\end{align}
for $i = 1, \ldots, n$, 
as long as $E(\norm{z_{i}}_{\mathcal{H}})$ and $\sigma^{2}$ is bounded from above, for $i=1, \ldots, n$,  where $\bvar=(\bvar_1, \cdots, \bvar_n)\trans$ are noise vector with mean zero and bounded variance and 
 \begin{equation}\label{effective_dimension}
 \gamma(\lambda) = \sum_{j=1}^{\infty} \frac{ \mu_j}{\mu_j + \lambda},\notag
 \end{equation}
is the effective dimension and $\{\mu_{j}\}_{j=1}^{\infty}$ are the eigenvalues of kernel $K$ used in $\hat{m}(\bx)$. 
\end{lemma}
The first term in \eqref{a1} denotes the order of bias term and the second term denotes the square root of the variance term. Specifically, we have the asymptotic mean square error for $\hat{m}$,
\begin{equation} 
       \mbox{AMSE}( \hat{m} ) = O(1) \times \left\{ \lambda \Norm{m}_{\mathcal{H}}^2 + n^{-1}\gamma(\lambda)  \right\}. 
       \label{amse}
     \end{equation}
For the $\ell$-th order of Sobolev space, we have $\mu_j \le C j^{-2 \ell}$ and 
    \begin{equation}
  \gamma({\lambda})  = \sum_{j=1}^{\infty} (1 + j^{2\ell}{\lambda})^{-1}  \leq  O\left(\lambda^{-1/(2\ell)}\right).\label{6} 
  \end{equation}
 Note that (\ref{amse}) is minimized when 
  $ \lambda  \asymp  \gamma(\lambda) /n, $ 
  which is equivalent to  ${\lambda} \asymp n^{-  2 \ell/(2\ell+1)}$ under (\ref{6}).  The  optimal rate $\lambda \asymp n^{- 2 \ell/(2\ell+1)}$ leads to 
          \begin{equation} 
       \mbox{AMSE}( \hat{m} ) =O (n^{-2\ell/(2 \ell+1)} )
       \label{amse2}
     \end{equation} 
     which is the optimal rate in Sobolev space, as discussed by \cite{stone1982}.      
     
To investigate the asymptotic properties of the kernel ridge regression imputation estimator, we express 
\begin{align}
\hat{\theta}_{I} &= \frac{1}{n} \sum_{i=1}^{n}\left\{ \delta_{i}y_{i} + (1-\delta_{i}) \hat{m}(\bx_{i}) \right\} \notag \\
&= \underbrace{\frac{1}{n}\sum_{i=1}^{n}m(\bx_{i})}_{R_{n}} + \underbrace{\frac{1}{n}\sum_{i=1}^{n}\delta_{i}\left\{   y_{i}  - m(\bx_{i}) \right\}}_{ S_{n} }  + \underbrace{\frac{1}{n}\sum_{i=1}^{n} (1-\delta_{i})\left\{ \hat{m}(\bx_{i}) - m(\bx_{i})  \right\}}_{T_{n}}.\notag
\end{align}
Therefore, as long as we show 
\begin{align}
T_{n} = \frac{1}{n} \sum_{i=1}^{n}\delta_{i}\left\{ \frac{1}{\pi(\bx_{i})} - 1  \right\}\left\{ y_{i} - m(\bx_{i})  \right\} + o_{p}(n^{-1/2}),\label{12} 
\end{align}
then we can establish the root-$n$ consistency. The following theorem formally states the theoretical result. A proof of Theorem 1 is presented in the supplementary material. 

\setcounter{theorem}{0} 

\begin{theorem}\label{main theorem}
Suppose Assumption \ref{A1}-\ref{A3} hold for a Sobolev kernel of order $\ell$, as long as 
\begin{equation} 
n\lambda \rightarrow 0,\quad n\lambda^{1/2\ell}\rightarrow \infty,
\label{con2} 
\end{equation} 
we have
$$n^{1/2} \left( \hat{\theta}_I - \theta \right) {\rightarrow}  N(0, \sigma^2 ) , 
$$
where
$$ \sigma^2 = \Var\{ E( Y \mid \bx) \} + E\{ \Var( Y \mid \bx)/\pi( \bx)   \} = \Var ( \eta)  $$ 
with 
\begin{align}\label{tilde_theta} 
\eta &=  m(\bx) + \delta\frac{1}{\pi(\bx)}  \left\{ y- m(\bx)\right\} . 
\end{align}
\end{theorem}

\begin{remark}
Note that the optimal rate ${\lambda} \asymp n^{- 2 \ell/(2\ell+1)}$ does not satisfy the first part of (\ref{con2}).  
 To control the bias part, we need a smaller $\lambda$ such as $\lambda =n^{-\kappa}$ with $\kappa > 1$. 
 Similar conditions are used for bandwidth selection for nonparametric kernel regression with bandwidth $h$: 
        $$ n h \rightarrow \infty \mbox{ and } n^{1/2}h^2 \rightarrow 0 $$
        for $\mbox{dim}(\bx)=1$. See \cite{wang09} for details. 
\end{remark}

\begin{remark} 
Theorem~\ref{main theorem} is presented for a Sololev kernel, and any kernel whose eigenvalues have the same tail behavior as Sobolev of order $\ell$ also has the result as Theorem~\ref{main theorem}.
For sub-Gaussian kernel whose eigenvalues satisfy that 
\begin{align}
    \mu_{j} \leq c_{1}\exp(-c_{2}j^{2}),\notag
\end{align}
where $c_{1}, c_{2}$ are positive constants, we can establish similar results. To see this, note that  
\begin{align}
    \gamma(\lambda) &= 
    \sum_{j=1}^{\infty} \frac{\mu_{j}}{\mu_{j} + \lambda    }\notag \\
    &\leq  c_{2}^{-1/2}  \{-\log(\lambda)  \}^{1/2} 
    + \frac{1}{\lambda} \int_{ c_{2}^{-1/2} \{- \log(\lambda)\}^{1/2} } \exp(-c_{2}z^{2})dz\notag\\
    &\leq c_{2}^{-1/2}  \{-\log(\lambda)  \}^{1/2} 
    + O(1),\notag
\end{align}
where the second term in the last equation can be obtained by the  Gaussian tail bound inequality. Therefore, as long as $n\lambda \rightarrow 0$ and $ n\{-\log(\lambda)\}^{-1/2}\rightarrow \infty$, we have $n^{-1}\bone_{n}\trans \ba = o_{p}(n^{-1/2})$ and the root-$n$ consistency can be established. 
\end{remark}

Note that the asymptotic variance of the imputation estimator is equal to $n^{-1} \sigma^2$, which is the lower bound of the semiparametric asymptotic variance discussed in \cite{robins94}. Thus, the kernel ridge regression imputation is asymptotically optimal. The main term (\ref{tilde_theta}) in the linearization in Theorem 1 is called the influence function \citep{hampel74}. The term influence function is motivated by
the fact that to the first order $\eta_i = m(\bx_{i}) + \delta_{i} \{ \pi(\bx_{i})\}^{-1}  \left\{ y_{i} - m(\bx_{i}) \right\}$ is the influence of a single observation on the estimator $\hat{\theta}_I$.

The  influence function  in (\ref{tilde_theta}) can be used for variance estimation of the KRR imputation estimator $\hat{\theta}_I$. The idea is to estimate the influence function 
$\eta_i = m(\bx_{i}) + \delta_{i} \{ \pi(\bx_{i})\}^{-1}  \left\{ y_{i} - m(\bx_{i}) \right\} $ and apply the standard variance estimator using $\hat{\eta}_i$. To estimate $\eta_i$, we need an estimator of $\pi(\bx)$. In the next section, we will consider a version of kernel ridge regression to estimate $\omega(x) = \{ \pi(\bx) \}^{-1}$ directly.  Once $\hat{\omega}_{i} (\bx)$ is obtained, we can use 
\begin{equation} 
\hat{\mbox{V}}  = \frac{1}{n} \frac{1}{n-1} \sum_{i=1}^n \left( \hat{\eta}_i - \bar{\eta}_n \right)^2  \notag
\end{equation} 
as a variance estimator of $\hat{\theta}_I$ in (\ref{1}), where 
\begin{equation} 
\hat{\eta}_i = \hat{m}(\bx_{i}) + \delta_{i} \hat{\omega}_{i} (\bx_i) \left\{ y_{i} - \hat{m}(\bx_{i})\right\}
\notag
\end{equation} 
and $\bar{\eta}_n = n^{-1} \sum_{i=1}^n \hat{\eta}_i$.

\section{Propensity score estimation}

 We now consider estimation of the propensity weight function $\omega(x) = \{ \pi(\bx) \}^{-1}$ using kernel ridge regression.   In order to estimate $\omega(\bx) = \{ \pi(\bx) \}^{-1}$, we wish to develop a nonparametric method 
 of estimating $\omega (\bx)$ using the same RKHS  theory. To do this,  we use the density ratio function estimation approach to propensity score function estimation proposed by \cite{wang2021}. To introduce the idea, we
 first define the following density ratio function 
\begin{equation} 
 g( \bx) = \frac{ f(\bx \mid  \delta =0 ) }{ f( \bx \mid  \delta = 1 ) },
\label{dr2}
\end{equation} 
and, by Bayes theorem, we have  
$$ \omega(\bx)= \frac{1}{ \pi(\bx) }  = 1+  c \cdot   g(\bx)
$$
where $c= \pr( \delta =0)/\pr(\delta =1)$. 
Thus, to estimate $\omega(\bx)$, we have only to estimate the density ration function $g(\bx)$ in (\ref{dr2}). 
Now, to estimate $g(\bx)$, we use the maximum entropy method \citep{nguyen2010} for  density ratio function estimation. \cite{kanamori2012statistical} also considered the M-estimator of the density ratio function with the Kullback-Leibler divergence.

For convenience, let $f_{k}(\bx) = f(\bx \mid  \delta = k )$, for $k = 0, 1$. To explain the M-estimation of $g(\bx)$,  note that $g(\bx)$ can be understood as the maximizer of the objective function on the right-hand-side of \eqref{LM} which is upper bounded by the Kullback-Leibler divergence between $f_{0}$ and $f_{1}$, i.e., 
 \begin{align} 
  D_{KL}(f_{0}, f_{1}) &= \max_{g > 0}   Q (g) + 1 \notag\\
  &=  \max_{g > 0} \int \log \left\{ g(\bx) \right\} f_{0}(\bx)  d \mu(\bx)   -   \int g (x) f_{1}( \bx)   d \mu(x)  + 1 \notag  \\
     &= \max_{g > 0} \int g(\bx) [\log \left\{ g(\bx) \right\} - 1] f_{1}(\bx)  d \mu(\bx) + 1.  \label{LM}
\end{align} 
That is, by \eqref{LM}, a sample version of $Q(g)$ can be written as 
\begin{align}
\hat{Q}(g) = \frac{1}{n_1} \sum_{i=1}^{n} \delta_{i} g(\bx_{i}) [\log \{ g(\bx_{i})\} - 1],\notag
\end{align}
where $n_{1} = \sum_{i=1}^{n} \delta_{i}$. 

 Since $g(\bx)$ is unknown, we want to impose constraints to formulate an M-estimation problem for $g(\bx)$. 
   Given $\hat{m}( \cdot)$, using the idea of model calibration \citep{wu2001}, 
   we would like to use 
$$ \frac{1}{n_1} \sum_{i=1}^n \delta_i g( \bx_i) \hat{m}(\bx_i) = \frac{1}{n_0} \sum_{i=1}^n (1-\delta_i) \hat{m}(\bx_i) 
$$
as a constraint for density ratio estimation, where $n_0=n-n_1$.  Note that it is algebraically equivalent to 
\begin{equation} 
 \frac{1}{n} \sum_{i=1}^n \delta_i \left\{ 1+ \frac{n_0}{n_1}  \cdot  {g}( \bx_i) \right\} \hat{m}(\bx_i) =  \frac{1}{n} \sum_{i=1}^n  \hat{m}(\bx_i) . \notag
\end{equation} 
        
Now, as we have $m\in\mathcal{H}$, and by the representer theorem in kernel ridge regression, we know that $\hat{m} \in \mbox{span}\{K(\cdot, \bx_{1}), \ldots, K(\cdot, \bx_{n})\}$. Thus, the calibration constraint is 
\begin{align}
\frac{1}{n_{1}}\sum_{i=1}^{n}\delta_{i}g(\bx_{i}) (K(\cdot, \bx_{1}), \ldots, K(\cdot, \bx_{n}   ))\trans = \frac{1}{n_{0}}\sum_{i=1}^{n}(1-\delta_{i}) (K(\cdot, \bx_{1}), \ldots, K(\cdot, \bx_{n}   ))\trans .  \label{LM_contraint}
\end{align}
This calibration property is also called  covariate-balancing property       \citep{imai2014covariate}. 
 Further, we want to incorporate with the normalization constraint  $\sum_{i=1}^{n} \delta_i \omega(\bx_{i})=n$, i.e., 
\begin{align}
\frac{1}{n_{1}}\sum_{i=1}^{n}\delta_{i}g(\bx_{i}) = \frac{1}{n_{0}}\sum_{i=1}^{n}(1-\delta_{i}). \label{LM_constraint_2}
\end{align}

Minimizing $\hat{Q}(g)$ subject to \eqref{LM_contraint} and \eqref{LM_constraint_2} is called the maximum entropy method. Using Lagrangian multiplier method,  the solution to this optimization problem can be written as 
\begin{align}
\log\{g(\bx)\} \equiv \log\{g(\bx; \bphi)\}  = \phi_{0} + \sum_{i=1}^{n}\phi_{i}K(\bx, \bx_{i}) 
\label{log-linear}
\end{align}
for some $\bphi = (\phi_{0}, \ldots, \phi_{n})\trans \in \mathbb{R}^{n+1}$. Thus, using the parametric form in \eqref{log-linear}, the optimization problem can be expressed as a dual form 
 \begin{equation} 
\hat{Q}_{0}(\bphi) =  \frac{1}{n_0} \sum_{i=1}^{n} (1-\delta_{i}) \log \{ g(\bx_i; \bphi)  \} - \frac{1}{n_1} \sum_{i=1}^{n} \delta_{i} g (\bx_i; \bphi),\notag
\end{equation}
to formulate a legitimate estimation of $g(\cdot)$. 
Further, define 
$h (\bx; \bphi_{s})= \log \{ g(\bx; \bphi)\} - \phi_{0}$, where $ \bphi_{s} = (\phi_{1},\ldots, \phi_{n})\trans$. In our  problem, to ensure the Representer theorem,  we wish to find $h$ that minimizes 
\begin{equation}
-\hat{Q}_{0}(g; \bphi)+ \tau \left\| h \right\|_{\mathcal{H} }^{2} \label{18} 
\end{equation} 
over $\bphi$. 

Hence, 
the solution to (\ref{18}) can be obtained as 
\begin{equation} \label{entropy_method}
\min_{ \bphi_{s} \in \mathbb{R}^{n}  } \left\{ 
\frac{1}{n_1} \sum_{i=1}^{n} \delta_{i} g (\bx_i; \bphi) - 
 \frac{1}{n_0} \sum_{i=1}^{n} (1-\delta_{i}) \log \{ g(\bx_i; \bphi)  \}
 + \tau \bphi\trans \bK  \bphi  \right\} 
\end{equation}
and $\phi_0$ is a normalizing constant satisfying 
\begin{equation}
 n_1 = \sum_{i=1}^{n} \delta_{i} \exp \{ \phi_0 + \sum_{j=1}^n  \hat{\phi}_j K( \bx_i, \bx_j ) \} .
 \label{phi0}
 \end{equation}

 Thus, we use 
\begin{equation}
 \hat{g} (x) = \exp \{ \hat{\phi}_0 + \sum_{j=1}^n  \hat{\phi}_j K( \bx, \bx_j)  \}
 \notag
 \end{equation} 
 as the maximum entropy estimator  of the density ratio function $g( \bx)$ using kernel method. Also, 
 \begin{equation} 
 \hat{\omega} (\bx)= 1+ \frac{n_0}{n_1} \hat{g}(\bx)\notag
 \end{equation} 
 is the maximum entropy estimator  of $\omega(x) = \{ \pi(\bx) \}^{-1}$.  The estimator of $\omega(x)$ satisfies the calibration property by construction.  That is, for any function $f(\bx) \in \mathcal{H}$, we have 
$$ n^{-1} \sum_{i=1}^n \delta_i \hat{\omega}(\bx_i) f(\bx_i) =  n^{-1} \sum_{i=1}^n  f(\bx_i).  
 $$



The tuning parameter $\tau$ is chosen to minimize
\begin{equation} 
D( \tau) = \left\| \frac{1}{n} \sum_{i=1}^n \delta_i \left\{ 1+ \frac{n_0}{n_1}\cdot \hat{g}_{\tau} ( x_i) \right\} \hat{m} ( x_i) - \frac{1}{n} \sum_{i=1}^n  \hat{m} ( x_i)  \right\|, 
\label{dr3}
\end{equation} 
where $\hat{m} ( x)$ is determined by kernel ridge regression estimation.  Thus, we can use the following two-step procedure to determine the tuning parameter $\tau$. 
\begin{description}

\item{[Step 1]} 
Use the kernel ridge regression to obtain $\hat{m} (\bx)$. 

\item{[Step 2]}  
Given $\hat{m} (\bx)$, find $\hat{\tau}$ that minimizes $D( \tau)$ in (\ref{dr3}). 
\end{description}


Further, we can also obtain the propensity score estimator based the above procedure, i.e.,
\begin{align}
\label{PS} 
\hat{\theta}_{{\rm PS}} = \frac{1}{n}\sum_{i=1}^{n}\delta_{i}\hat{\omega}(\bx_{i})y_{i}.
\end{align}
We now establish the root-$n$ consistency of the propensity score estimator in the following theorem.  
 \begin{theorem}\label{P3T2}
 Under regularity conditions stated in the supplementary material, we have 
 \begin{align}
 n^{1/2} \left( \hat{\theta}_{{\rm PS}} - \theta\right)  
\rightarrow N(0,\sigma^{2}),
 \end{align}
 where $\sigma^{2} = \Var(\eta)$ and 
 \begin{align}
     \eta = m(\bx) + \delta \frac{1}{\pi(\bx)} \{y - m(\bx)\}. \notag
 \end{align}
 \end{theorem}
 
 Theorem \ref{P3T2} implies that the propensity score estimator in (\ref{PS}) using the above procedure is asymptotically equivalent to the KRR imputation estimator and  achieves the same asymptotic variance as the KRR imputation estimator. 
 The regularity conditions and a sketched proof of Theorem \ref{P3T2} are presented in the supplementary material. We can use a linearized variance estimator to get a valid variance estimate based on Theorem \ref{P3T2}, similar to Theorem \ref{main theorem}.

\begin{remark} 
As the objective function in (\ref{entropy_method}) is convex, we apply the limited-memory Broyden-Fletcher-Goldfarb-Shanno algorithm to solve  the optimization problem with the following first order partial derivatives: 
\begin{align}
  \frac{\partial U }{\partial \phi_{0}} = &   \frac{1}{n_{1}}\sum_{i=1}^{n}\delta_{i}\exp\left( \phi_{0} + \sum_{j=1}^{n}\phi_{j}K(\bx_{i}, \bx_{j})  \right) - 1,\notag\\   \frac{\partial U }{\partial \phi_{k}} = & \frac{1}{n_{1}}\sum_{i=1}^{n}\delta_{i}K(\bx_{i}, \bx_{k})\exp\left( \phi_{0} + \sum_{j=1}^{n}\phi_{j}K(\bx_{i}, \bx_{j})  \right)    - \frac{1}{n_{0}}\sum_{i=1}^{n}(1-\delta_{i})K(\bx_{i}, \bx_{k}) \notag\\   &+ 2  \tau\sum_{i=1}^{n}K(\bx_{i}, \bx_{k})\phi_{i},\quad k = 1, \ldots, n, \notag
\end{align}
where $U$ is the objective function in (\ref{entropy_method}). \end{remark}

\section{Simulation Study}



To compare with the existing methods and to evaluate the finite-sample performance of the proposed imputation method and its variance estimator, we conduct a limited simulation study. In this simulation, we consider the continuous study variable with three different data generating models.   In the three models, we keep the response rate around $60\%$ and $\var(Y) \approx 10$. Also, $\bx_{i} =  (x_{i1}, x_{i2}, x_{i3}, x_{i4})\trans$ are generated independently {element-wise} from the {uniform} distribution on the support $(1,3)$. In the first model A, we use a linear regression model 
$
y_{i} = 3 + 2.5x_{i1} + 2.75 x_{i2} +  2.5 x_{i3} + 2.25 x_{i4} + \sigma\epsilon_{i}$
to obtain $y_i$, 
where $\{\epsilon_{i}\}_{i=1}^{n}$ are generated from standard normal distribution and $\sigma = 3^{1/2}$. In the model B, we use $
y_{i} = 3 + (1/35)x_{i1}^{2}x_{i2}^{3}x_{i3} +  0.1x_{i4} + \sigma\epsilon_{i}$
to generate data with a {nonlinear} structure. The model C for generating the study variable is 
$y_{i} = 3 + (1/180)x_{i1}^{2}x_{i2}^{3}x_{i3}x_{i4}^{2} + \sigma\epsilon_{i}.$

In addition to $\{(\bx_i, y_i), i = 1, \ldots, n\}$, we consider two response mechanisms. The response indicator variable $\delta$'s for each mechanism are independently generated from  different {Bernoulli} distributions. 
In the first response mechanism, the probability for the {Bernoulli} distribution is $\logit(\bx_i\trans \bbeta + 2.5)$, where  ${\bbeta} = (-1.1, 0.5, -0.25, -0.1)\trans$ and $\mbox{logit}(p) = \log\{p / (1-p)\}$. In the second response  mechanism, the probability for the {Bernoulli} distribution is $\logit(-0.3  + 0.7 x_{1}^{2}  - 0.5 x_{2} - 0.25 x_{3} - 0.25 x_{4})$. We considered two sample sizes  $n = 500$ and $n = 1,000$. 

The reproducing kernel Hilbert space we employed in the simulation study is the second-order Sobolev space. In particular, we used tensor product RKHS to extend a one-dimensional Sobolev space to the  multidimensional space. From each sample, we consider four imputation methods: imputation  and propensity score methods related to kernel ridge regression and the others are B-spline and linear regression.
{For the B-spline method, we employ the generalized additive model by R package `mgcv'. Specifically, we used cubic spine with 15 knots for each coordinate with restricted maximum likelihood estimation method.}
We used $B=1,000$ Monte Carlo samples in the simulation study.

\begin{figure}
\begin{center}
\includegraphics[width=1.0\textwidth]{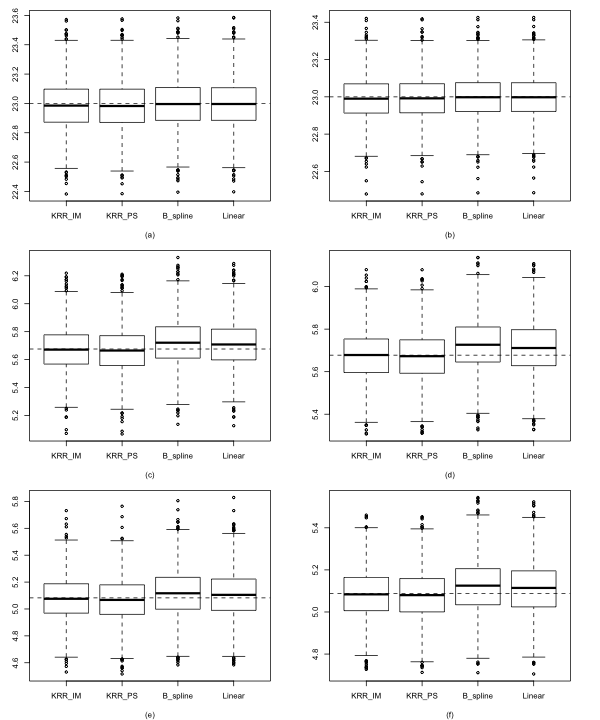}
\end{center}
\caption{
Boxplots with four estimators for model A ((a) for $n=500$ and (b) for $n = 1000$), model B ((c) for $n=500$ and (d) for $n = 1000$) and model C ((e) for $n=500$ and (f) for $n = 1000$) under first response  mechanism with true values (dashes). KRR\_IM, kernel ridge regression imputation estimator; KRR\_PS, kernel ridge regression propensity score estimator.
}
\label{fig1} 
\end{figure}

\begin{figure}
\begin{center}
\includegraphics[width=1.0\textwidth]{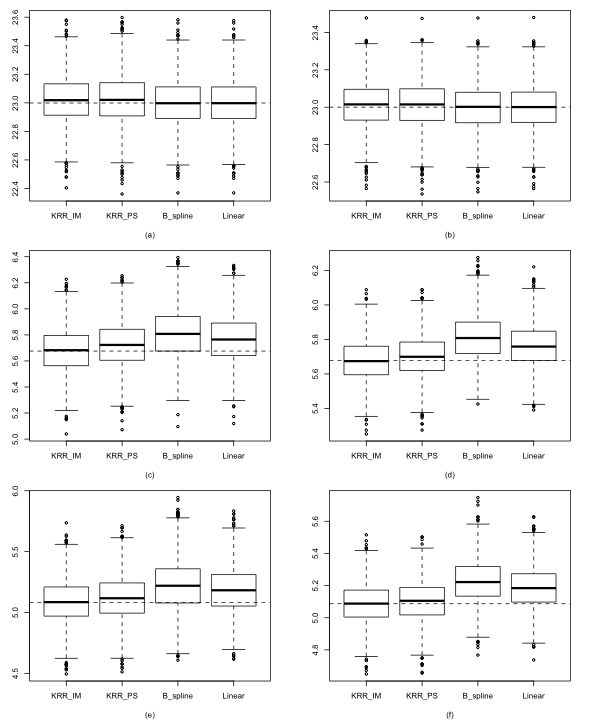}
\end{center}
\caption{
Boxplots with four estimators for model A ((a) for $n=500$ and (b) for $n = 1000$), model B ((c) for $n=500$ and (d) for $n = 1000$) and model C ((e) for $n=500$ and (f) for $n = 1000$) under second response  mechanism with true values (dashes). KRR\_IM, kernel ridge regression imputation estimator; KRR\_PS, kernel ridge regression propensity score estimator.
}
\label{fig2} 
\end{figure}

The simulation results of the four point estimators 
for the first response mechanism and for the second response mechanism are summarized in Figure \ref{fig1} and Figure \ref{fig2}, respectively.   
The simulation results in  Figure \ref{fig1} and Figure \ref{fig2} show that  four methods show similar results under the linear model (model A), but both kernel ridge regression imputation estimators and propensity score estimators show robust performance  under the nonlinear models (models B and C). All kernel ridge regression  related methods provide negligible biases in all scenarios. 

In addition, we have computed the proposed variance estimators under kernel ridge regression imputation with the corresponding kernel. 
In Table \ref{MAR, linear, KRR}, the relative biases (in percentage) of the proposed variance estimator  and the coverage rates of two interval estimators under $90\%$ and $95\%$ nominal coverage rates are presented. 
The relative bias of the variance estimator are relatively low, which confirms the validity of the proposed variance estimator. Furthermore, the interval estimators show good performances in terms of the coverage rates. 


\begin{table}
\caption{Relative biases (R.B.)  of the proposed variance estimator, coverage rates (C.R.) of the $90\%$ and $95\%$ confidence intervals for imputed estimators and propensity score estimators under kernel ridge regression with second-order Sobolev kernel and Gaussian kernel for continuous responses (KRR\_IM, kernel ridge regression imputation estimator; KRR\_PS, kernel ridge regression propensity score estimator) 
}\label{MAR, linear, KRR}
\begin{tabular}{cccccccccc}
  \hline
  & & \multicolumn{4}{c}{First Missing Mechanism} & \multicolumn{4}{c}{Second Missing Mechanism} \\
   &  & \multicolumn{2}{c}{KRR\_IM} & \multicolumn{2}{c}{KRR\_PS} & \multicolumn{2}{c}{KRR\_IM} & \multicolumn{2}{c}{KRR\_PS} \\
 \multirow{-3}{*}{ Model}   & \multirow{-3}{*}{Criteria}  &   n=500 & n=1000 &   n=500 & n=1000  &   n=500 & n=1000 &   n=500 & n=1000 \\ 
  \hline
    &  R.B(\%) & 0.09 & -2.80 & 0.15 & -3.14 & 3.40 & 2.74 & -1.68 & -1.90 \\ 
    A&  C.R.(90\%) & 90.30 & 89.95 & 90.30 & 89.75 & 90.25 & 90.60 & 89.15 & 89.85 \\ 
    &  C.R.(95\%) & 95.50 & 94.95 & 95.70 & 95.00 & 95.20 & 95.45 & 94.65 & 94.80 \\ 
    \\
    \cline{2-10}
    &  R.B(\%) & -2.77 & -5.42 & -5.77 & -6.60 & -6.07 & -3.42 & -11.25 & -6.23 \\ 
    B&  C.R.(90\%) & 89.55 & 89.70 & 89.20 & 89.20 & 88.05 & 90.05 & 87.75 & 89.30 \\ 
    &  C.R.(95\%) & 94.25 & 94.55 & 93.85 & 94.10 & 94.15 & 94.70 & 93.35 & 94.10 \\ 
    \\
    \cline{2-10}
    &  R.B(\%) & -7.43 & -3.97 & -12.24 & -6.22 & -9.38 & -2.29 & -13.62 & -4.34 \\ 
    C&  C.R.(90\%) & 87.95 & 88.70 & 86.70 & 88.75 & 88.80 & 89.50 & 87.50 & 89.75 \\ 
    &  C.R.(95\%) & 93.35 & 94.20 & 92.35 & 93.70 & 93.95 & 95.15 & 93.25 & 94.70\\
   \hline
\end{tabular}
\end{table}

\section{Application}

We applied the kernel ridge regression with the kernel of second-order Sobolev space to study the $\mbox{PM}_{2.5}(\mu g/m^{3})$ concentration measured in Beijing, China \citep{liang2015assessing}. Hourly weather conditions: temperature, air pressure, cumulative wind speed, cumulative hours of snow and cumulative hours of rain are available from 2011 to 2015. Meanwhile, the averaged sensor response is subject to missingness. In December 2012, the missing rate of $\mbox{PM}_{2.5}$ is relatively high with missing rate $17.47\%$. We are interested in estimating the mean $\mbox{PM}_{2.5}$ in December with both imputed and propensity score kernel ridge regression estimates. The point estimates and their 95\% confidence intervals are presented in the Table \ref{CI_Table}. 
 As a benchmark, the confidence interval computed from complete cases and confidence intervals for the imputed estimator under linear model \citep{kim2009unified}  are also presented there. 

\begin{table}[htb]
\centering
\caption{Point estimates (P.E.), standard error (S.E.) and $95\%$ confidence intervals (C.I.) for imputed mean $\mbox{PM}_{2.5}$ in December, 2012 under kernel ridge regression (KRR\_IM, kernel ridge regression imputation estimator; KRR\_PS, kernel ridge regression propensity score estimator.)
}\label{CI_Table}
\begin{tabular}{c|ccc}
  \hline
   Estimator & P.E. & S.E. & $95\%$ C.I. \\ 
  \hline
 Complete & 109.20 & 3.91 & (101.53, 116.87) \\ 
  Linear  & 99.61 & 3.68 & (92.39, 106.83) \\ 
  KRR\_IM & 101.92 & 3.50 & (95.06, 108.79) \\ 
  KRR\_PS & 102.25 & 3.50 & (95.39, 109.12) \\ 
   \hline
\end{tabular}
\end{table}


As we can see, the performances of {kernel ridge regression imputation estimators are similar} and created narrower $95\%$ confidence intervals. Furthermore,  the imputed $\mbox{PM}_{2.5}$ concentration during the missing period 
is relatively lower than  the fully observed weather conditions on average.  Therefore, if we only utilize the complete cases to estimate the  mean of $\mbox{PM}_{2.5}$, the severeness of air pollution would be over-estimated.

\section{Concluding Remarks}
 We consider kernel ridge regression  as a tool for nonparametric imputation and propensity score weighting. 
    The proposed kernel ridge regression imputation can be used as a general tool for nonparametric imputation. By choosing different kernel functions, different  nonparametric imputation methods can be developed. Asymptotic properties of the propensity score estimator are also established. The unified theory developed in this paper  enables us to make valid nonparametric statistical inferences about the population means 
    under missing data. 
 

There are several possible extensions of the research. First, the theory can be extended to other nonparametric imputation methods, such as smoothing splines \citep{claeskens2009}, thin plate spline \citep{wahba1990spline}, Gaussian process regression \citep{rasmussen2005},  or deep kernel learning \citep{bohn2019}. The theoretical results in this paper can be used as building-blocks for  establishing the statistical properties  of these sophisticated nonparametric imputation methods. Second, instead of using ridge-type penalty term, one can also consider other penalty functions such as the smoothly clipped absolute deviation penalty \citep{FL2001} or adaptive lasso \citep{zou2006}. Such penalty functions can be potentially useful for handling high dimensional covariate problems.  Also, the proposed method can be used for causal inference, including estimation of average treatment effect from observational studies (\citealp{morgan2014counterfactuals};  \citealp{yangding2020}).  Developing tools for causal inference using the kernel ridge regression-based propensity score method will be an important extension of this research. 

\bibliographystyle{chicago}

\bibliography{ref2}

\end{document}